\documentclass[jgr]{agutex}

\usepackage{color}
\usepackage{url}
\usepackage{lineno}
\usepackage{graphicx}
\usepackage{multirow}

\authorrunninghead{WANG ET AL.} \titlerunninghead{CME Deflection}

\begin{document}

\title{On the Propagation of a Geoeffective Coronal Mass Ejection during March 15 -- 17, 2015}

\author{Yuming Wang,\altaffilmark{1,7,*} Quanhao Zhang,\altaffilmark{1,8} Jiajia Liu,\altaffilmark{1,8}
Chenglong Shen,\altaffilmark{1,7,9} Fang Shen,\altaffilmark{2} Zicai Yang,\altaffilmark{2}
T. Zic,\altaffilmark{3} B. Vrsnak,\altaffilmark{3} D. F. Webb,\altaffilmark{4}
Rui Liu,\altaffilmark{1,8,9} S. Wang,\altaffilmark{1,8} Jie Zhang,\altaffilmark{5}
Qiang Hu,\altaffilmark{6} and Bin Zhuang\altaffilmark{1,8}}

\affil{$^1$ CAS Key Laboratory of Geospace Environment,
Department of Geophysics and Planetary Sciences, University of Science and
Technology of China, Hefei, Anhui 230026, China}
\affil{$^2$ State Key Laboratory of Space Weather, National Space Science Center, CAS, Beijing 100190, China}
\affil{$^3$ Hvar Observatory, Faculty of Geodesy, Kaciceva 26, HR-10000 Zagreb, Croatia}
\affil{$^4$ ISR, Boston College, Newton, MA 02459, USA}
\affil{$^5$ Department of Physics and Astronomy,
George Mason University, MSN 6A2, Fairfax, VA 22030, USA}
\affil{$^6$ Department of Space Science and CSPAR, The University of Alabama in Huntsville, Huntsville, Alabama, USA}
\affil{$^7$ Synergetic Innovation Center of Quantum Information and Quantum Physics,
University of Science and Technology of China, Hefei 230026, China}
\affil{$^8$ Collaborative Innovation Center of Astronautical Science and Technology, Hefei 230026, China}
\affil{$^9$ Mengcheng National Geophysical Observatory, University of Science and Technology of China, Hefei 230026, China}
\affil{$^*$ Corresponding Author, Contact: ymwang@ustc.edu.cn}

\begin{abstract}
The largest geomagnetic storm so far in the solar cycle 24 was produced by a fast coronal mass ejection (CME) originating on 2015
March 15. It was an initially west-oriented CME and expected to only cause a weak geomagnetic disturbance.
Why did this CME finally cause such a large geomagnetic storm? We try to find some clues by investigating its propagation
from the Sun to 1 AU. First, we reconstruct the CME's kinematic properties in the corona from the SOHO and SDO imaging data
with the aid of the graduated cylindrical shell (GCS) model. It is suggested that the CME propagated to the west
$\sim$$33^\circ$$\pm$$10^\circ$ away from the Sun-Earth
line with a speed of about 817 km s$^{-1}$ before leaving the field of view of the SOHO/LASCO C3 camera.
A magnetic cloud (MC) corresponding to this CME was measured in-situ by the Wind spacecraft two days after the CME
left LASCO's field of view. By applying two
MC reconstruction methods, we infer the configuration of the MC as well as some kinematic information, which implies
that the CME possibly experienced an eastward deflection on its way to 1 AU.
However, due to the lack of observations from the STEREO spacecraft, the CME's kinematic evolution in interplanetary
space is not clear. In order to fill this gap, we utilize numerical MHD simulation, drag-based CME propagation model
(DBM) and the model for CME deflection
in interplanetary space (DIPS) to
recover the propagation process, especially the trajectory, of the CME from $30 R_S$ to 1 AU under the constraints
of the derived CME's kinematics near the Sun and at 1 AU. It is suggested that the trajectory of the CME was deflected
toward the Earth by about $12^\circ$, consistent with the implication from the MC reconstruction at 1 AU.
This eastward deflection probably contributed to the CME's unexpected geoeffectiveness by pushing
the center of the initially west-oriented CME closer to the Earth.
\end{abstract}

\begin{article}

\section{Introduction}
As the most important driver of severe space weather, coronal mass ejections (CMEs) and their
geoeffectiveness have been studied intensively. Previous statistical studies have shown that
not all the front-side halo CMEs are geoeffective~\citep[e.g.,][]{Webb_etal_2001, Wang_etal_2002a, Zhao_Webb_2003,
Yermolaev_etal_2005}, and not all
non-recurrent geomagnetic storms can be tracked back to a CME~\citep[e.g.,][]{Cane_etal_2000,
Cane_Richardson_2003, Yermolaev_etal_2005, Zhang_etal_2007}. These phenomena may cause some
failed predictions of the geoeffectiveness of CMEs. The recent notable event exhibiting such
a failure was on 2015 March 15 when a fast CME originated from the west hemisphere.
Space Weather Prediction Center (SWPC) of NOAA initially forecasted that the CME would at most
cause a very minor geomagnetic disturbance labeled as G1, a scale used by SWPC to measure the intensity
of geomagnetic storms. However, the CME produced the largest geomagnetic storm so far,
at G4 level with the provisional $Dst$ value of $-223$ nT, in the current solar cycle
24~\citep{Kataoka_etal_2015, Wood_etal_2016}. The major geomagnetic storm was called the ``2015
St. Patrick's Day" event as its main phase and peak occurred on March 17, and the surprising CME
was selected as a campaign event by International Study of Earth-affecting Solar Transients
(ISEST)\footnote{\url{http://solar.gmu.edu/heliophysics/index.php/03/17/2015_04:00:00_UTC}},
a program under SCOSTEP, and also by Coupling, Energetics and Dynamics of Atmospheric Regions
(CEDAR)\footnote{\url{http://cedarweb.vsp.ucar.edu/wiki/index.php/2015_Workshop:The_March_17_2015_great_storm}}.

\begin{figure*}[b]
  \centering
  \includegraphics[width=\hsize]{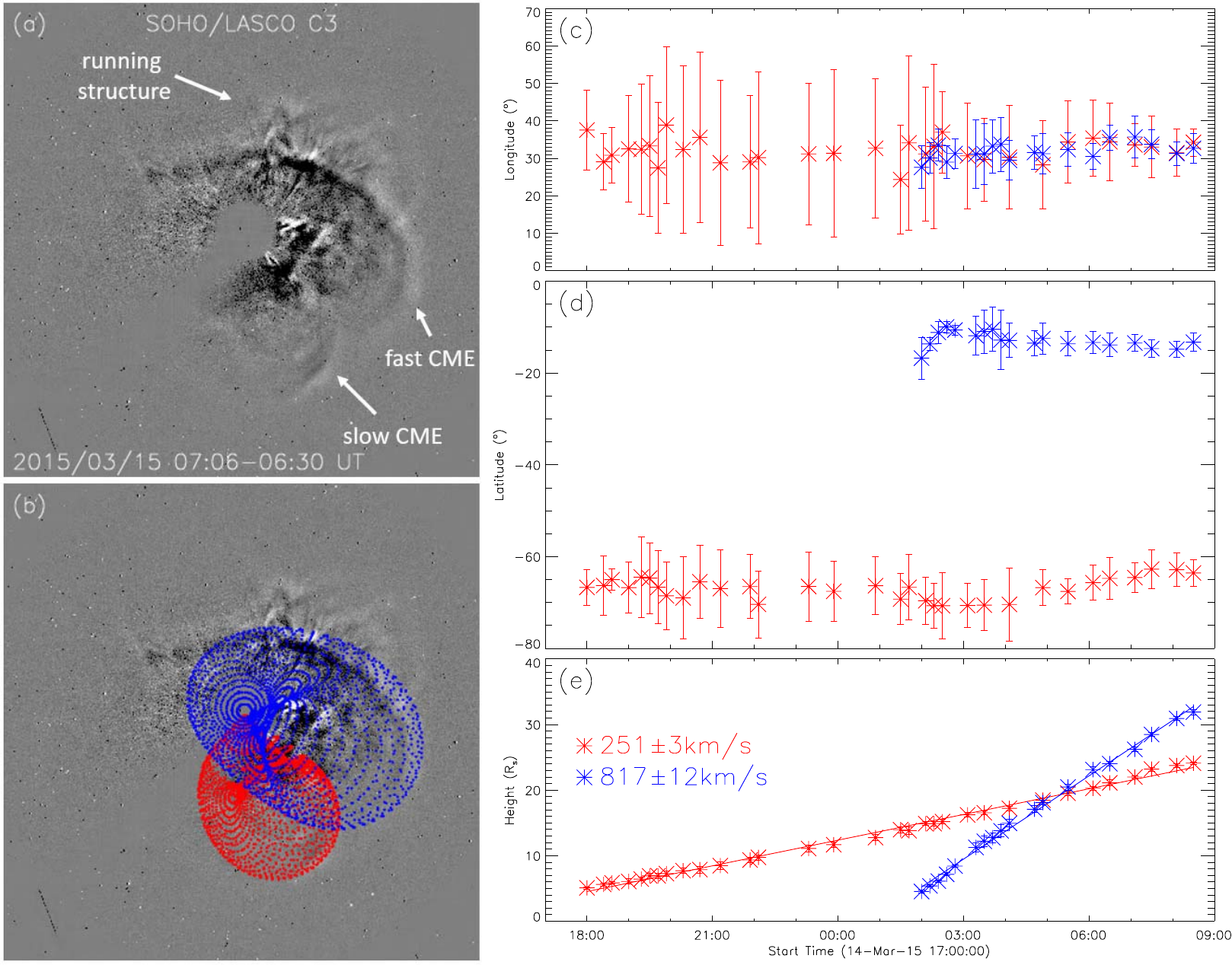}
  \caption{(a)--(b) SOHO/LASCO C3 difference images showing the fast CME (blue) as well as the preceding slow CME (red)
with the GCS fitting meshes superimposed. (c)--(e) Longitudes, latitudes and heights of the leading edges
of the two CMEs obtained from the GCS fitting. The line in Panel (g) is the linear fit to the CME height assuming
a reasonable uncertainty of $\pm1 R_S$.}\label{fg_gcs}
\end{figure*}

Such an unexpected phenomenon naturally raises the first question for the forecasting of the
geoeffectiveness of a CME, i.e., whether or not a CME will hit the Earth even though we know
the source location and initial kinematic properties of the CME. A full understanding of the propagation trajectory of a
CME from the Sun to 1 AU is the key to this question.
Of course, it is not the only factor determining the geoeffectiveness
of a CME. The magnetic field strength and the orientation of the CME flux rope, which directly affect
the strength and duration of the interval of the south-component of the magnetic field,
is also important for determining its geoeffectiveness.

It has been well accepted that the CME's trajectory can be deflected in the corona (within a few tens
of solar radii, $R_S$). \citet{Wang_etal_2011}
illustrated that such deflections can be classified into three types: asymmetrical expansion,
non-radial ejection and deflected propagation. In a statistical sense, CMEs
tend to be deflected toward the equator during solar minimum~\citep[e.g.,][]{MacQueen_etal_1986,
Cremades_Bothmer_2004, Wang_etal_2011} or
deflected away from coronal holes~\citep[e.g.,][]{Gopalswamy_etal_2003, Gopalswamy_etal_2009,
Cremades_etal_2006}. The physics behind these deflections is that the gradient of the
magnetic energy density may cause the CME to move toward
the place where the magnetic energy density reaches the minimum, usually the location
of the heliospheric current sheet~\citep{Shen_etal_2011, Gui_etal_2011, Zuccarello_etal_2012,
Isavnin_etal_2013, Kay_etal_2013}. Although the CME's deflection
in the corona could be tens of degrees and may change the geoeffectiveness of a CME, it still can
be monitored by coronagraphs~\citep[e.g.,][]{Mostl_etal_2015}. Thus, the possible deflection of a CME in interplanetary space rather
than the deflection in the corona is one of the major sources of uncertainty in the prediction of the
CME impact at the Earth.

The possibility of the CME deflection in interplanetary space was first proposed
by~\citet{Wang_etal_2004b}. They suggested that, different from the deflection in the corona,
the CME's trajectory in interplanetary space could be deflected due to the velocity difference between the CME and the
ambient solar wind. For a fast CME, the solar wind plasma and interplanetary magnetic field will be
piled up from the west and ahead of it, leading to a net deflection force toward the east; for
a slow CME, the picture is the opposite. A kinematic model (called DIPS,
{\it Deflection in InterPlanetary Space}, hereafter) was therefore developed~\citep{Wang_etal_2004b}.
Such deflections are thought to be gradual and much
slower than that in the corona, but the total amount of the deflection angle is comparable to
that in the corona as it takes place over a much longer distance. Such evidence can also be found in previous
studies~\citep[e.g.,][]{Wang_etal_2002a, Wang_etal_2004b, Wang_etal_2006a, Kilpua_etal_2009, Lugaz_etal_2010,
Isavnin_etal_2014, Kay_Opher_2015}. One of the most comprehensive analysis
of the CME's trajectory in interplanetary space was done by~\citet{Wang_etal_2014}
for a slow CME, which was proven to experience a westward
deflection all the way from the corona to 1 AU with a total deflection angle of more than 20 degrees.

For the 2015 March 15 CME, there were no STEREO~\citep[Solar TErrestrial RElations Observatory,][]{Kaiser_etal_2008}
data as the twin spacecraft were behind the Sun and not taking images. All the
information of the CME came from the remote-sensing data provided by the Solar and Heliospheric Observatory (SOHO)
and the in-situ data by Wind (or ACE) spacecraft at 1 AU. The interplanetary space
between the corona to 1 AU thus had an observational gap, and therefore the propagation of the CME from the Sun
to 1 AU is unclear. In this paper, we try to recover the kinematic evolution of the CME from the limited
observations and fill the gap with the aid of models. We particularly focus on the trajectory of the
CME to demonstrate how the CME behavior in interplanetary space favors its strong geoeffectiveness.

\section{Kinematics of the CME in the corona}
The fast CME of interest
first appeared in the field of view (FOV) of LASCO~\citep[Large Angle
and Spectrometric Coronagraph,][]{Brueckner_etal_1995} C2 camera on board the SOHO on March 15 at about 01:36 UT,
and left the FOV of LASCO/C3, which monitors the corona within 30 $R_S$, around 09 UT.
It was a partial halo CME with most material ejected toward the west as shown in Figure~\ref{fg_gcs}a.

\begin{figure*}[tb]
  \centering
  \includegraphics[width=\hsize]{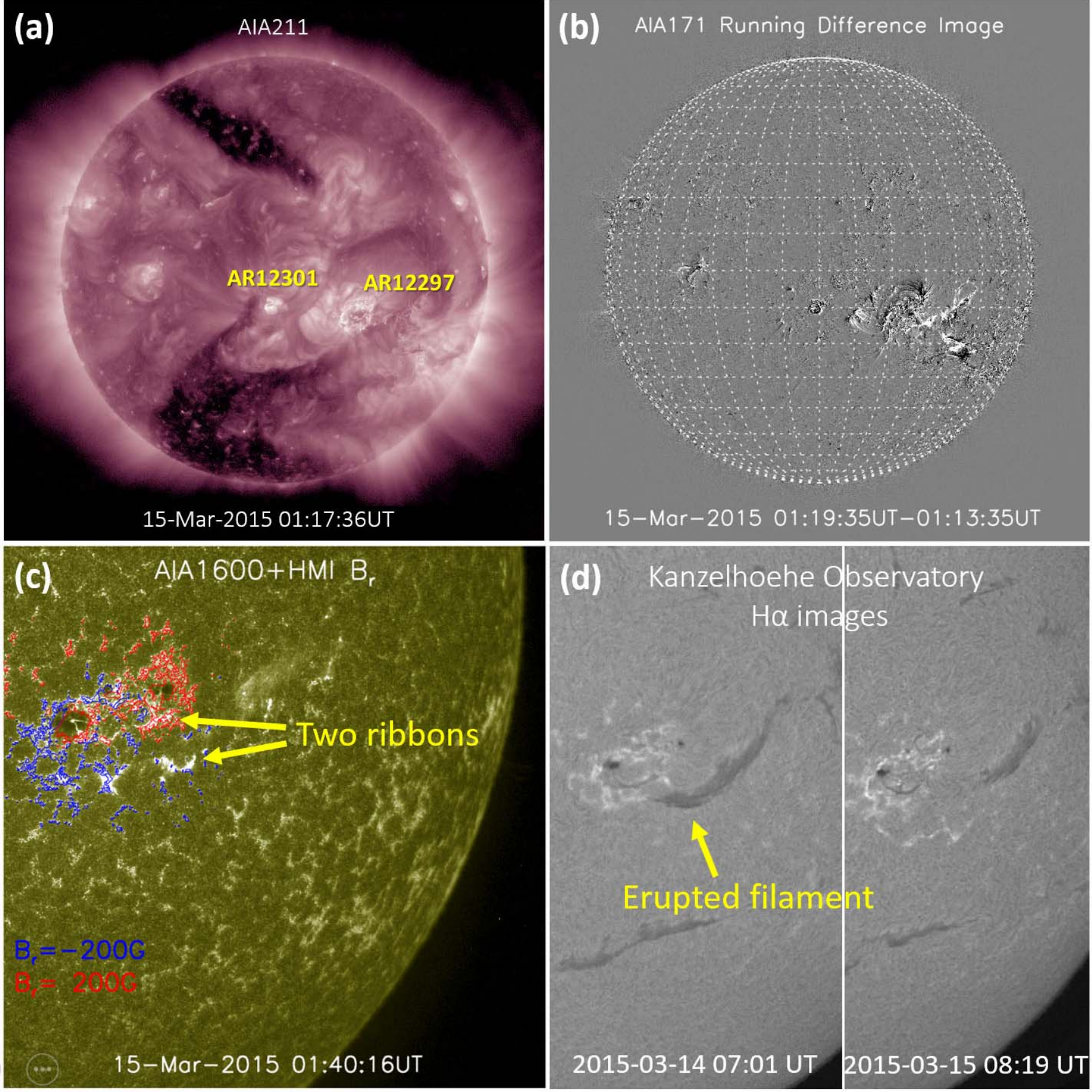}
  \caption{(a) and (b) SDO/AIA 211\AA\ image and 171\AA\ difference image, respectively, showing the launch site of the fast CME, which
  is in the AR 12297 with a large south-polar CH extending to the south-east of the AR.
(c) SDO/AIA 1600\AA\ image combined with the contours of the radial component of the photospheric magnetic field from the Helioseismic and Magnetic Imager~\citep[HMI,][]{Hoeksema_etal_2014} on board SDO,
showing the associated two-ribbon flare and the magnetic topology of the AR. (d) Two daily $H\alpha$ images from Kanzelhoehe Observatory,
showing the partial disappearance of the associated filament.}\label{fg_source}
\end{figure*}

By examining the solar EUV images, e.g., Figure~\ref{fg_source}a and \ref{fg_source}b, taken
by the Atmospheric Imaging Assembly~\citep[AIA,][]{Lemen_etal_2012} on board the Solar Dynamics Observatory (SDO),
we can identify that the source region
of the CME is active region (AR) 12297 on the west hemisphere of the Sun with a large coronal hole (CH)
to the south-east of it. The eruption took place around the location of
W35S15 with filament-like material moving toward the south-west direction. Meanwhile, two bright ribbons can be seen
in the AIA 1600\AA\ passband (Fig.\ref{fg_source}c), suggesting a typical eruptive flare. The two
daily $H\alpha$ images taken by Kanzelhoehe Observatory before and after the eruption (Fig.\ref{fg_source}d)
did show a disappearance of a segment of a thick filament near the AR. However, as will be discussed below,
there was another slow CME perhaps originating from the same AR between the times of the two $H\alpha$ images. Thus,
the association of the disappeared filament segment in the $H\alpha$ images to the CME is unclear. The AIA
EUV images suggest a clear filament eruption during the fast CME, and therefore it is very likely that the disappeared filament is
associated with this CME. Considering that the filament
is a good tracer of the CME flux rope, one may estimate that the tilt angle, i.e., the angle between
the main axis of the CME flux rope and the solar equator, is initially within the range of $\pm 20^\circ$.

It should be noted that the coronal conditions during the fast CME traveling through the LASCO FOVs
were complicated. The movie (see supplementary material) made from the SOHO/LASCO images (or see the example image Fig.\ref{fg_gcs}a)
reveals that (1) there was a structure running faster than the fast CME
along the north-west direction, and (2) there was a slow CME, which was launched earlier on March 14, propagating
toward the south. The running structure appeared in the LASCO FOV at the same time as the fast CME appeared. It not only
ran faster than but also looked fainter than the fast CME. This running structure might be the shock wave
driven by the fast CME, or an independent magnetic structure that erupted from the Sun. We incline to the latter because
the running structure was much faster than the fast CME and its shape was much narrow than and different from the CME's front. However,
near the eruption of the fast CME, we cannot find any other notable eruption signatures on the visible solar disk.
Although stealth CMEs exist, they tend to be slow~\citep[e.g.,][]{Robbrecht_etal_2009, Wang_etal_2011, HowardT_Harrison_2013}.
Thus, the fast running structure was probably a real ejection from the backside of the Sun.
Since the structure propagated faster than and ahead of the CME of interest, we do not consider any possible interaction
between them.

For the preceding slow CME, there are two possibilities of its source locations (due to the lack of STEREO observations).
One was suggested by
Gopalswamy and Yashiro at the ISEST workshop in October, 2015. They thought that it was a backside
event because the pre-existing streamer disturbed by the slow CME was moving to the south pole, suggesting that its location
was on the backside. The other possibility is that it was a front-side CME. There was a notable eruptive signature
in the same AR 12297 around 12:00 UT on March 14. The time and location match well with the appearance and the speed of
the slow CME in the LASCO/C2 FOV. If the first possibility was true, there should be no interaction between
the slow CME and the fast CME of interest. But if the second possibility was true, the two CMEs might have interacted with each other.

With the aid of
foward modeling, e.g., the GCS model~\citep{Thernisien_etal_2009, Thernisien_2011}, we then analyze the kinematics
of the two CMEs as well as this possible interaction. Since SOHO/LASCO provides only one angle of view, the GCS fitting
suffers from a larger uncertainty than that when the STEREO data are available. Thus, we
reduce the degrees of freedom during the fitting by setting three free parameters, the tilt angle, aspect ratio and angular width,
to be constant, and only vary the other three free parameters, the longitude, latitude and the height.
Thanks to a sufficient number of images in the time sequence, these free parameters of the GCS model can still
be roughly constrained by trial and error. The uncertainties in these parameters are estimated by following the
method of~\citet{Thernisien_etal_2009}, i.e., by decreasing the goodness-of-fit between the leading edge of the
CME and the model by 10\%. Since the leading edge of the CME in all the images is determined manually by hand-clicks,
which may increase the errors, the uncertainties inferred by the above method are underestimated.
For the fast CME, the best value of the tilt angle is about $-22^\circ$, the aspect ratio about 0.66,
and the angular width about $83^\circ$ (edge-on) or $172^\circ$ (face-on). For the preceding slow CME,
they are $-20^\circ$, 0.47, $56^\circ$ and $97^\circ$, respectively. The uncertainties in the tilt angle, aspect ratio
and the angular width are about $20^\circ$, 0.12 and $30^\circ-50^\circ$. These uncertainties are very large,
and thus these fitting values are only used for reference.

For the other three time-dependent parameters, the fitting results are shown in
Figure~\ref{fg_gcs}c--\ref{fg_gcs}e. The uncertainty in the longitude is about $10^\circ$, that in the latitude
is better, about $5^\circ$, and that in the height is less than one solar radius. The two CMEs
almost propagated along the same longitude, which is around $30^\circ$, but in latitude, the two CMEs were
separated by about $50^\circ$. Considering the angular width of the two CMEs, they might marginally interact
with each other if the preceding slow CME was a front-side event. According to the height-time plot shown
in Figure~\ref{fg_gcs}e, the two CMEs traveled through the LASCO FOVs at a speed of 817 and 251 km s$^{-1}$, respectively.
The leading edge of the fast CME caught up with the leading edge of the preceding slow one around 05:20 UT.
Thus, the possible interaction should start around 03:00 UT and end before the fast CME
left the FOV of LASCO/C3. However, neither considerable acceleration of the preceding slow CME nor deceleration of
the fast CME can be found in the height-time profiles.
An interesting phenomenon is that the preceding slow CME was systematically deflected toward lower latitude,
and the fast CME slightly toward higher latitude. Due to the significant uncertainties, we cannot conclude
that the two CMEs interacted based on such small deflections. Nevertheless, we can conclude that
the two CMEs at most experienced a very weak interaction, which has little influence on the kinematics of the
fast CME in interplanetary space. Note, there is also a great possibility that the two
CMEs did not interact at all, because the preceding one might come from the backside of the solar disk.

The inferred kinematic evolution of the two CMEs in the corona suggest that
the fast CME should be able to encounter the Earth but the slow CME is unlikely
to encounter the Earth.
Since the fast CME may drive a shock at front, what we fitted with GCS model
is probably not the leading edge of its flux rope but the driven shock. The recent work by \citet{Good_Forsyth_2016}
suggested that the longitudinal extent of the flux rope carried by a CME is typically about $60^\circ$,
smaller than its driven shock if any. Thus, the angular width of the fast CME derived from the GCS model
is probably overestimated. If this were the case, the fast CME's flux rope might just graze the Earth with
a weaker geoeffectiveness.

\begin{figure*}[tb]
  \centering
  \includegraphics[width=0.8\hsize]{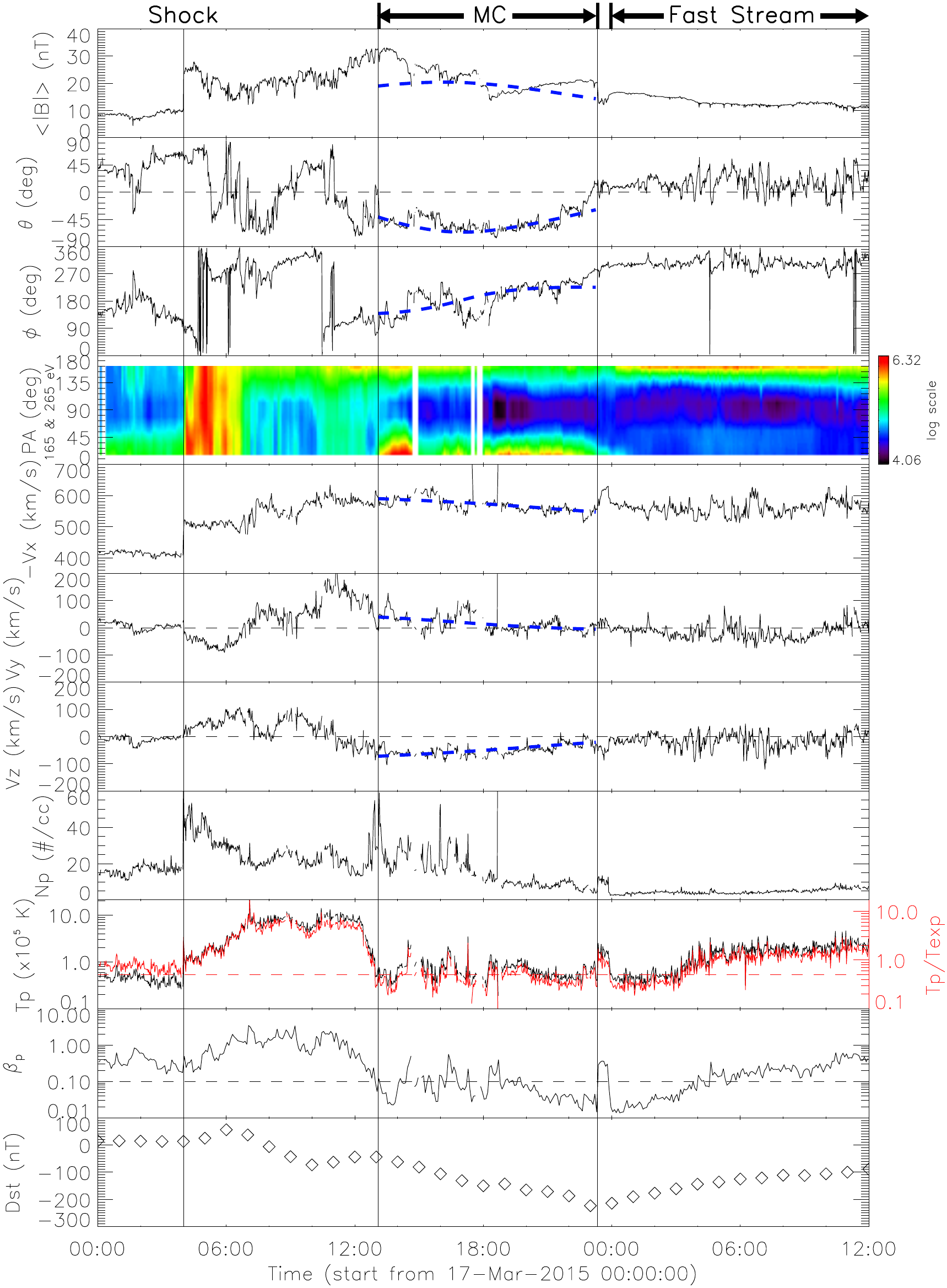}
  \caption{In-situ measurements from Wind spacecraft (the first 10 panels) and the provisional $Dst$ index from the
WDC for Geomagnetism, Kyoto Dst index service (the last panel).
The magnetic field and its elevation and azimuthal angles in GSE coordinates measured by Wind/MFI are presented
in the first three panels, the pitch angle of suprathermal electrons of 165 and 265 eV measured by Wind/3DP is shown in the fourth panel,
the three components of solar wind velocity in GSE coordinates, number density, temperature and $\beta$ of protons measured by Wind/SWE
are displayed in the next 6 panels. The red line in the ninth panel is the ratio of
measured proton temperature to the expected temperature, which is calculated based on the empirical formula by~\citep{Lopez_Freeman_1986}.
The dashed blue curves is the fitting of the velocity-modified cylindrical force-free flux rope model.}\label{fg_wind}
\end{figure*}

\section{In-situ observations at 1 AU}
Two days later, a fast forward shock followed by a magnetic cloud (MC) was recorded by the Wind
spacecraft~\citep{Lepping_etal_1995, Ogilvie_etal_1995, Lin_etal_1995} as
shown in Figure~\ref{fg_wind}. The shock arrived at 1 AU at 04:00 UT on March 17. Its driver, the MC,
occurred between 13:05 and 23:20 UT, and is characterised by the reduced fluctuation in the magnetic field, the large and
smooth rotation of the magnetic field direction, evident bi-directional suprathermal electron beams and the low
temperature and proton $\beta$. According to the classification proposed by \citet{Bothmer_Schwenn_1998} and
\citet{Mulligan_etal_1998}, the MC is an ESW-type flux rope. A fast stream, probably from the CH (as seen in Fig.\ref{fg_source}a),
was catching up with the MC, causing an interaction region during 23:20 -- 23:55 UT, when the density
and temperature were enhanced, the magnetic field strength reduced and the magnetic field rotation ceased.
Our identification of the MC is the same as that by~\citet{Kataoka_etal_2015}. The MC and the shock sheath ahead
of it caused a double-peak major geomagnetic storm with the provisional $Dst$ value of $-223$ nT (see the last panel of Fig.\ref{fg_wind}).

The MC is the interplanetary counterpart of
the CME originating on March 15 because of the following two reasons. (1) If the MC corresponds to the CME, the transit
time of the CME from its first appearance in the LASCO/C2 and the arrival at 1 AU is about 59.5 hours, and the
average transit speed is about 690 km s$^{-1}$, which is consistent with the CME speed, 817 km s$^{-1}$, in the FOV of
LASCO/C3 and the measured MC speed, $\sim600$ km s$^{-1}$, at 1 AU.
(2) Except for the preceding slow CME mentioned in the last section, there was no other CME candidate
during March 14 -- 16 responsible for the MC according to the LASCO observations. The slow CME propagated
far away from the ecliptic plane and was probably on the other side of the Sun. Thus, it should not be detected
near the Earth. It should be noted that~\citet{LiuY_etal_2015}
also analyzed this event and proposed a different scenario that there were two interplanetary CMEs (ICMEs),
corresponding to the slow and fast CMEs on March 14 and 15, respectively, and the main ICME, i.e., the one
corresponding to the fast CME, was identified in the interval from about 18 UT on March 17 to 16 UT on the next day.
This interpretation
is not in agreement with our above analysis. The Wind data shown in Figure~\ref{fg_wind} reveal that the
smooth rotation of magnetic field vector and the bi-directional electron streams ceased at the end of March 17.

There are various models developed to reconstruct MCs from one-dimensional in-situ data, including
cylindrically symmetrical force-free flux rope models~\citep[e.g.,][]{Goldstein_1983,
Marubashi_1986, Burlaga_1988, Lepping_etal_1990, Wang_etal_2015}, asymmetrically cylindrical (non)force-free flux rope
models~\citep[e.g.,][]{Mulligan_Russell_2001, Hu_Sonnerup_2002, Hidalgo_etal_2002a, Cid_etal_2002,
Vandas_Romashets_2003} and torus-shaped flux rope models~\citep[e.g.,][]{Romashets_Vandas_2003,
Marubashi_Lepping_2007, Hidalgo_Nieves-Chinchilla_2012}.
Here we use the velocity-modified cylindrical force-free flux rope (VFR) model,
which considers the propagation and expansion of a MC as well as the plasma poloidal motion inside
the MC~\citep{Wang_etal_2015}, to fit the MC. In this model, the magnetic field is described by the
Lundquist solution, and the velocity
is incorporated under the assumptions of self-similar evolution and magnetic flux conservation.
This model is proven to yield similar results as the cylindrically symmetrical force-free flux rope
model by \citet{Lepping_etal_1990}. We choose this model because the fitting results contain some kinematic information of the CME, which
may provide some clues on the trajectory of the CME in interplanetary space.
In addition, we also apply the Grad-Shafranov (GS) reconstruction
technique~\citep{Hu_Sonnerup_2002} to the MC to see the similarity and difference between the model results.

\begin{table*}[tb]
\begin{center}
\footnotesize
\caption{Fitting values of the free parameters of the velocity-modified cylindrical force-free flux rope model}\label{tb_par}
\begin{tabular}{c|ccccccccccc}
\hline
Model &$B_0$ & $R_{MC}$ & $\theta$ & $\phi$ & $H$ & $d$     & $v_x$ & $v_y$ & $v_z$ & $v_e$ & $v_p$\\
(1) & (2) & (3) & (4) & (5) & (6) & (7) & (8) & (9) & (10) & (11) & (12) \\
\hline
VFR& 32    & 0.09     & $-45^\circ$ & $348^\circ$ & $+1$ & $-0.82R_{MC}$ & $-540$ & 59 & $-27$ & 51 & 45 \\
GS& $\sim23$  & $\sim0.05$  & $-47^\circ$ & $281^\circ$ & $+1$ & $-0.013AU$ & / & / & / & / & / \\
\hline
\end{tabular}
From the left to the right, the columns give (1) the model used to derive the parameters, (2)
the magnetic field strength at the MC's axis in units of nT,
(3) the radius of the MC in units of AU, (4) the elevation angle of the MC's axis in the GSE coordinates,
(5) the azimuthal angle of the MC's axis in the GSE coordinates, (6) the handedness of the MC, (7) the
closest approach of the observational path to the MC's axis, (8--10) the propagation velocity of the MC
in the GSE coordinates in units of km s$^{-1}$, (11) the expansion speed of the MC in units of km s$^{-1}$
and (12) the poloidal speed of the MC plasma in units of km s$^{-1}$.
\end{center}
\end{table*}

The best-fit values of the free parameters of the VFR model for the MC of interest are given
in Table~\ref{tb_par}, and the fitting curves are plotted as the dashed blue lines in Figure~\ref{fg_wind}, which match the observed
profiles fairly well in both magnetic field and velocity.
In particular, we highlight the following parameters:
(i) Sign of the helicity or handedness of the MC is $+1$, which obeys the pattern that the southern hemisphere of
the Sun usually accumulates positive helicity~\citep[e.g.,][]{Rust_Kumar_1996}. (ii) Closest approach is
$0.82$ $R_{MC}$, where $R_{MC}$ is the
radius of the MC, indicating that the observational path is far away from the MC's axis. (iii) Orientation of the MC's axis is
$\theta=-45^\circ$ and $\phi=348^\circ$ in GSE coordinates, corresponding to a tilt angle of about $-46^\circ$
when projected on the plane-of-the-sky. Considering that the magnetic polarity in the CME source region is negative/positive on
the southern/northern side of the filament associated with the CME (see Fig.\ref{fg_source}c) and the helicity of the MC
is positive, this value is close to the CME's tilt angle of $-22^\circ$ estimated
from the GCS fitting. The orientation also suggests that the angle between the axis of the observed portion of the MC and
the Sun-Earth line is
about $50^\circ$, meaning that the flank of the MC was passed through~\citep[e.g.,][]{Janvier_etal_2013}.
(iv) Combination of (i) and (iii) gives how the magnetic field lines wind in the MC,
which is in agreement with the distribution of the magnetic field polarities in the
CME source region, i.e., positive/negative polarity region on the north-west/south-east side of the associated filament
(Fig.\ref{fg_source}c).

\begin{figure*}[tb]
  \centering
  \includegraphics[width=\hsize]{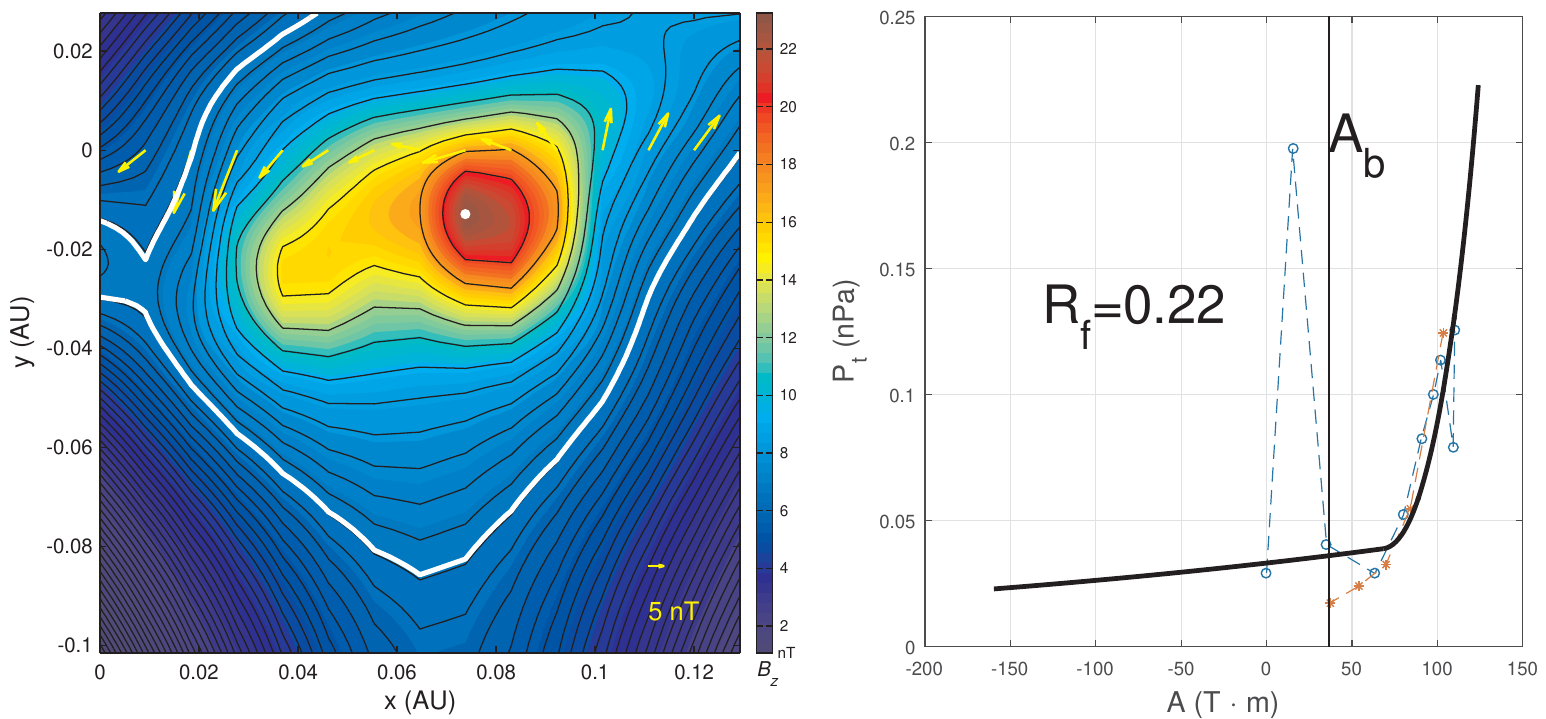}
  \caption{The GS reconstruction results. Left panel:  the cross-sectional map, contours of the magnetic flux function $A$
and the axial field $B_z$ in color as indicated by the color bar, of the cylindrical flux rope. The yellow arrows along $y=0$
represent measured transverse field along the spacecraft path. White dot denotes the center of the flux rope of maximum axial
field strength. Right panel:  the corresponding measured transverse pressure $P_t=B_z^2/2\mu_0+p$, the sum of the axial
magnetic pressure and the plasma pressure, versus $A$ along the spacecraft path. A functional fit of $P_t(A)$ is shown by
the thick black curve with a corresponding fitting residue $R_f$ denoted. The vertical line of $A=A_b$
marks the boundary of flux rope which is also highlighted by the thick white contour line in the left panel.
Refer to \citet{Hu_etal_2004} for more details of the GS reconstruction technique.}\label{fg_gs}
\end{figure*}

The results of the GS reconstruction are shown in Figure~\ref{fg_gs} and the fitting parameters are listed in
Table~\ref{tb_par} too.
It is a substantially different technique from the VFR fitting which is a type of forward models.
The GS reconstruction neither presets the shape of the flux rope nor assumes a fore-free state, but uses the GS equation to infer the
two and a half dimensional distribution of magnetic field at the cross-section of the flux rope from the observed
magnetic field and thermal pressure along the spacecraft path under the assumption of being time-stationary and magnetohydrostatic.
The fitting of $P_t(A)$, the sum of the axial magnetic pressure and the thermal pressure,
which is essential to the GS reconstruction, yields a residue $R_f=0.22$~\citep{Hu_etal_2004}
as a measure of goodness of fit. As judged from the right panel of Figure~\ref{fg_gs}, the interpretation
of a flux-rope configuration shown in the left panel is valid for the limited region within the white contour
line and under the assumption that significant axial current exists at the center. The GS reconstruction gives the
same handedness as VFR model, and we can read from Figure~\ref{fg_gs} that
the magnetic field at the center is about 23 nT and the radius about 0.08 AU in $y$-axis or 0.05 AU in $x$-axis, slightly
smaller than but comparable to those derived from the VFR model. The closest approach is about $-0.013$ AU.
For the orientation, the GS model is quite consistent with the VFR model in the elevation angle,
but deviates significantly in the azimuthal angle. The angle between the
orientations of the GS model and the VFR model is about $-67^\circ$, the largest inconsistency between the two models.
The comparison suggests that the solution of the fit to the in-situ measurements is not unique.
The most sensitive parameter is the orientation of the flux rope axis. Besides, the selection of the
boundaries of a MC might also significantly affect the fitting results (private communication with K. Marubashi and Q. Hu).
It is difficult to evaluate which one is more reliable. We list the two
possible solutions here for reference and also to raise the question for further attention.

\citet{Riley_etal_2004}
performed `blind tests' by applying five different fitting techniques, including the cylindrical linear force-free flux
rope model, the elliptical cross-section non-force-free flux rope model and the GS model,
to a MHD simulated MC. The largest deviation among these model results is in the orientation,
especially when the observational path is far away from the MC's axis. The March 17 MC
encountered the Earth with the closest approach of $0.82$ $R_{MC}$, falling into this scenario. Thus, it is
not surprising that we get quite different results in the orientation from the different models.
However, the tests by \citet{Riley_etal_2004} do suggest that the fitting
technique based on a cylindrical force-free flux rope is a useful tool.

\section{Inferring the CME trajectory in interplanetary space}

The in-situ measurements of the solar wind velocity reveal that there were significant components of the velocity
in $+y$ and $-z$ directions in the GSE coordinates. The VFR model suggests that
the $y$-component of the propagation velocity of the MC is $\sim59$ km s$^{-1}$,
and the $z$-component is about $-27$ km s$^{-1}$. According to the tilt angle of the CME, the part of the CME
 closer to the Sun-Earth line was above the ecliptic plane, and therefore the $y$- and $z$-component velocities do imply that
the CME was approaching the Sun-Earth line and the ecliptic plane. Since the $z$-component velocity is smaller
than the $y$-component velocity and the longitude is more important than the latitude in this case, we only consider
the $y$-component velocity in the following analysis.

\begin{figure*}[tb]
  \centering
  \includegraphics[width=\hsize]{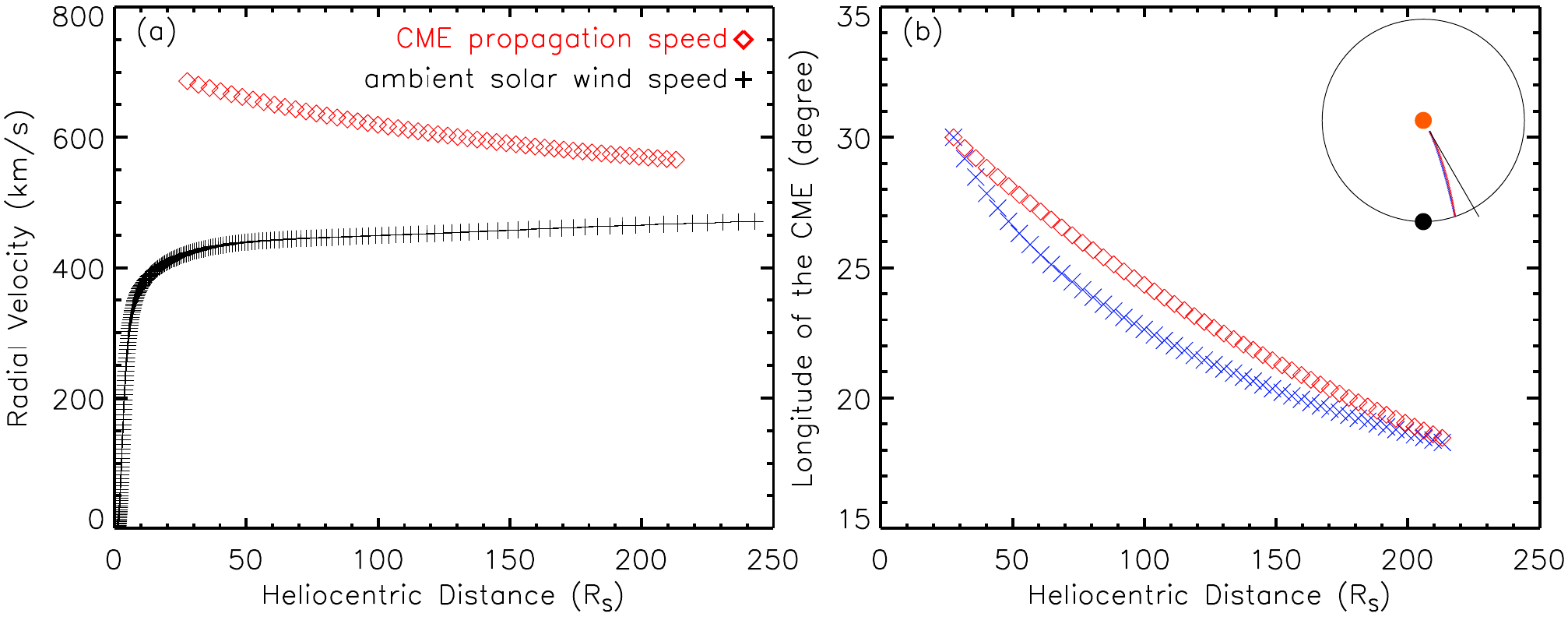}
  \caption{(a) The ambient solar wind speed obtained
from the 3D MHD simulation (the black + signs) and the CME
propagation speed derived from DBM model (the red diamonds). (b) The predicted CME longitude by DIPS model (the red
diamonds) and the estimated change of the CME longitude based on in-situ data (the blue crosses). The inset
at the upper-right corner shows the CME trajectory predicted by the DIPS model (the red line) and by the MC fitting (the blue line,
almost completely overlapped with the red line) in the ecliptic plane, in which
the Sun is denoted by the orange dot, the Earth by the black dot, and the radial direction (the straight black line) is
plotted for comparison.}\label{fg_mc_top}
\end{figure*}

It was shown by \citet{Wang_etal_2015} in the statistical study of 72 MCs that the
propagation velocity perpendicular to the radial direction is 19 km s$^{-1}$ on average, and only 8\% of
the events had a perpendicular velocity larger than 60 km s$^{-1}$. Thus, the $y$-component propagation velocity
in this case is significant enough to indicate an eastward deflection of the MC from the radial direction.
By comparing to the radial propagation speed, we may infer that the
deflection rate in the ecliptic plane, $\frac{v_y}{v_x}$, is about $-0.1$.
Assuming that the CME kept the deflection rate all the way from $30 R_S$, where the CME left the FOV of the
LASCO/C3 (Fig.\ref{fg_gcs}g), to 1 AU, the deflection angle can be calculated by the
equation 17 in~\citet{Wang_etal_2015}, i.e., $\Delta\Phi=\frac{v_y}{v_x}\ln\frac{1 AU}{30 R_S}\approx-12^\circ$.
The change of the CME's propagation direction estimated by this method is shown as the blue crosses in Figure~\ref{fg_mc_top}b.
Such an eastward deflection made the path of the Earth cutting through the CME closer to the CME center, and
therefore enhanced the geoeffectiveness of the initially west-oriented CME.

It is interesting to think about the cause of the possible eastward deflection of the CME. First, the fast
stream following the CME clearly modified the MC structure by, e.g., increasing the strength of
the magnetic field in the rear portion of the MC, but we think that it probably is not the major cause of the eastward deflection.
This CME is a structure embedded in a co-rotating interaction region (CIR)
characterized by the fast stream behind the CME and the slow stream ahead of it. One may imagine that the overtaking fast
stream would push the CME toward the west. The solar image, e.g., Fig.~\ref{fg_source}a, also suggests that the CME
should be deflected toward the west because of the presence of the large CH on the south-east of it. However, it is
probably not the case as implied by the in-situ data and the VFR model results. Thus, we think
that the CME's trajectory was actually controlled by the slow stream ahead of it.
The velocity difference between the CME and the preceding slow stream is about 10 times (roughly estimated by eyes from the in-situ data)
of that between the CME and the following fast stream.
Based on the DIPS picture~\citep{Wang_etal_2004b, Wang_etal_2014}, such velocity differences may lead to a more significant
pile-up of solar wind plasma and magnetic field ahead or on the west of the CME than that behind or on the east of it, and
therefore cause an eastward deflection.

Now we use the DIPS model to infer the CME trajectory with the constraints of the solar and in-situ
observations. To run the DIPS model, we need the CME propagation speed and
the ambient solar wind speed, both of which are the functions of the heliocentric distance.
As in \citet{Wang_etal_2014}, the ambient solar wind speed is obtained from the 3-dimensional (3D)
MHD simulation~\citep{Feng_etal_2003, Feng_etal_2005, ShenF_etal_2009, ShenF_etal_2011a}
for Carrington Rotation 2161 which covered the period from February 29
to March 27. We choose the simulated solar wind speed along the latitude of $-11^\circ$ and
within the longitude of $0^\circ$--$50^\circ$, along which the CME propagated, to generate the averaged
solar wind speed (the black plus signs in Fig.\ref{fg_mc_top}a).

The CME propagation speed is derived by the drag-based model~\citep[DBM,][]{Vrsnak_etal_2010, Vrsnak_etal_2013},
starting at the distance of $r_0=30 R_S$ with the initial speed of the CME leading edge $v_0=817$ km s$^{-1}$.
The simulated solar wind speed mentioned above is adopted.
We adjust the drag coefficient $\Gamma$
to match the DBM model output, the arrival time of the MC and the speed, with the in-situ observations at 1 AU.
After several trials, the best value of $\Gamma$ is found to be about $0.13\times10^{-7}$ km$^{-1}$, with which
the modeled MC arrival time is 13:32 UT and the speed of the MC leading edge is 622 km s$^{-1}$ at 1 AU,
close to the observed arrival time 13:05 UT and measured speed of about 600 km s$^{-1}$.
It should be noted that the speed of the CME leading edge consists of two components: the
propagation speed and the expansion speed. The VFR model has suggested that
the expansion speed is approximated to be one tenth of the CME radial propagation
speed. The ratio of the expansion speed to the radial propagation speed will hold as long as
the CME self-similarly evolved with a constant angular width.
By deducting the expansion speed from the speed of the CME leading edge, we obtain the
CME propagation speed as the function of the distance (see red diamonds in Fig.\ref{fg_mc_top}a).

The red diamonds in Figure~\ref{fg_mc_top}b show the trajectory of the CME derived from the DIPS model.
It is suggested that, from $30 R_S$ to 1 AU, the CME was deflected by about $12^\circ$ toward the east,
in agreement with the deflection angle, $-12^\circ$, estimated by the VFR model based on the in-situ observations.
It is difficult to evaluate the error of the predicted deflection angle. The direct error comes from the
uncertainties in both the CME and solar wind speeds. By considering an uncertainty of $\pm10\%$ in them,
the predicted deflection angle of this CME is about ${-12^\circ}_{-7^\circ}^{+8^\circ}$. However, since DIPS is a kinematic
model, the error may also come from other unknown factors which are not included in the model.
In the previous study by~\citet{Wang_etal_2014}, the deflection angle of the slow CME originating on 2008 September 12
is predicted as ${8^\circ}_{-9^\circ}^{+12^\circ}$, which is much smaller than the expected value, $\sim30^\circ$,
derived from the observations. Thus the same problem is also applicable to the fast CME investigated here.

\section{Summary}
Here, we study the fast CME originating on 2015 March 15, which caused the largest geomagnetic storm
so far in solar cycle 24. With the aid of forward modeling, the kinematic properties of the CME
in the corona are obtained based on the solar imaging data. Within 30 $R_S$ the CME propagated west of the Sun-Earth
line along the longitude of about $30^\circ$ and the latitude of about $15^\circ$.
The propagation speed is about 817 km s$^{-1}$. Its angular width is wide enough to make its east
flank overlap the Sun-Earth line. On the other hand, the in-situ data at 1 AU suggests the flank of a MC arriving
at the Earth. Two different models are applied to the MC to infer the configuration
of the MC. It is found that the handedness is consistent but
the orientation of the flux rope is quite different between the models. As suggested
by \citet{Riley_etal_2004}, such a large deviation in the orientation is probably due to
the spacecraft being too far away from the MC's axis. Currently, it is still difficult to evaluate the
reliability of the model results.

Due to the lack of the STEREO observations, the propagation of the 2015 March CME in the heliosphere is unclear.
We then try to recover the interplanetary evolution process of the CME from the information at the two ends: near the Sun and at 1 AU.
The VFR model results based on the in-situ data suggest that the CME experienced
a significant eastward deflection with the ratio of $\frac{v_y}{v_x}$ of about $-0.1$, implying a $12^\circ$-deflection
toward the Earth. The detailed trajectory of the CME between the two ends
are further reconstructed by using the numerical simulation (for background solar wind), DBM model (for the CME
propagation speed) and DIPS model (for the CME trajectory) under the constraints of the CME's kinematics obtained from the
solar and in-situ observations. The reconstructed trajectory is bent toward the Earth, quite consistent with the deflection
implied by the VFR model. This eastward deflection pushed the center of the initially west-oriented
CME closer to the Earth and probably contributed to the unexpected strong geoeffectiveness of the CME.
However, the lack of interplanetary observations causes that the above inference for this case
cannot be fully validated though it sounds reasonable, and the origin of this strong geomagnetic storm is still somewhat mysterious.

Some models applied in this study can be run and tested online.
One can go to \url{http://space.ustc.edu.cn/dreams/} for the DIPS model and the VFR model,
and to \url{http://oh.geof.unizg.hr/DBM/dbm.php} for the DBM model.

\begin{acknowledgments}
We acknowledge the use of the data from SOHO, SDO
and Wind spacecraft and the $H\alpha$ images from Kanzelhoehe Observatory.
SOHO is a mission of international cooperation between ESA and NASA, and
SDO is a mission of NASA's Living With a Star Program.
We also acknowledge the discussion of the event with T. Berger, N. Gopalswamy, P. Hess, Y. Liu,
K. Marubashi, C. Moestl, T. Rollett, M. Temmer, C.-C. Wu and S. Yashiro, under the ISEST program.
We thank the anonymous referees for their constructive comments. This work is supported by grants from NSFC (41131065, 41421063),
CAS (Key Research Program KZZD-EW-01 and 100-Talent Program), and
the fundamental research funds for the central universities. Y.W. is also supported by NSFC 41574165,
C.S. by NSFC 41274173, F.S. and Z.Y. by NSFC 41474152 and MOST 973 key project 2012CB825601, and R.L. by NSFC 41222031.
B.V. and T.Z. acknowledge the financial support by
Croatian Science Foundation under the project 6212 ``Solar and Stellar Variability".

\end{acknowledgments}




\begin{thebibliography}{66}
\providecommand{\natexlab}[1]{#1}
\expandafter\ifx\csname urlstyle\endcsname\relax
  \providecommand{\doi}[1]{doi:\discretionary{}{}{}#1}\else
  \providecommand{\doi}{doi:\discretionary{}{}{}\begingroup
  \urlstyle{rm}\Url}\fi

\bibitem[{\textit{Bothmer and Schwenn}(1998)}]{Bothmer_Schwenn_1998}
Bothmer, V., and R.~Schwenn, The structure and origin of magnetic clouds in the
  solar wind, \textit{Ann. Geophys.}, \textit{16}, 1, 1998.

\bibitem[{\textit{Brueckner et~al.}(1995)}]{Brueckner_etal_1995}
Brueckner, G.~E., et~al., The large angle spectroscopic coronagraph ({LASCO}),
  \textit{Sol. Phys.}, \textit{162}, 357--402, 1995.

\bibitem[{\textit{Burlaga}(1988)}]{Burlaga_1988}
Burlaga, L.~F., Magnetic clouds and force-free field with constant alpha,
  \textit{J. Geophys. Res.}, \textit{93}, 7217, 1988.

\bibitem[{\textit{Cane and Richardson}(2003)}]{Cane_Richardson_2003}
Cane, H.~V., and I.~G. Richardson, Interplanetary coronal mass ejections in the
  near-earth solar wind during 1996--2002, \textit{J. Geophys. Res.},
  \textit{108(A4)}, 1156, {doi:10.1029/2002JA009,817}, 2003.

\bibitem[{\textit{Cane et~al.}(2000)\textit{Cane, Richardson, and {St.
  Cyr}}}]{Cane_etal_2000}
Cane, H.~V., I.~G. Richardson, and O.~C. {St. Cyr}, Coronal mass ejections,
  interplanetary ejecta and geomagnetic storms, \textit{Geophys. Res. Lett.},
  \textit{27}, 3591--3594, 2000.

\bibitem[{\textit{Cid et~al.}(2002)\textit{Cid, Hidalgo, Nieves-Chinchilla,
  Sequeiros, and Vi\~{n}as}}]{Cid_etal_2002}
Cid, C., M.~A. Hidalgo, T.~Nieves-Chinchilla, J.~Sequeiros, and A.~F.
  Vi\~{n}as, Plasma and magnetic field inside magnetic clouds: a global study,
  \textit{Sol. Phys.}, \textit{207}, 187--198, 2002.

\bibitem[{\textit{Cremades and Bothmer}(2004)}]{Cremades_Bothmer_2004}
Cremades, H., and V.~Bothmer, On the three-dimensional configuration of coronal
  mass ejections, \textit{Astron. \& Astrophys.}, \textit{422}, 307--322, 2004.

\bibitem[{\textit{Cremades et~al.}(2006)\textit{Cremades, Bothmer, and
  Tripathi}}]{Cremades_etal_2006}
Cremades, H., V.~Bothmer, and D.~Tripathi, Properties of structured coronal
  mass ejections in solar cycle 23, \textit{Adv. in Space Res.}, \textit{38},
  461--465, 2006.

\bibitem[{\textit{Feng et~al.}(2003)\textit{Feng, Wu, Wei, and
  Fan}}]{Feng_etal_2003}
Feng, X., S.~T. Wu, F.~Wei, and Q.~Fan, A class of {TVD} type combined
  numerical scheme for {MHD} equations with a survey about numerical methods in
  solar wind simulations, \textit{Space Sci. Rev.}, \textit{107}, 43--53, 2003.

\bibitem[{\textit{Feng et~al.}(2005)\textit{Feng, Xiang, Zhong, and
  Fan}}]{Feng_etal_2005}
Feng, X., C.~Xiang, D.~Zhong, and Q.~Fan, A comparative study on 3-d solar wind
  structure observed by {Ulysses} and {MHD} simulation, \textit{Chinese Sci.
  Bull.}, \textit{50}, 672--678, 2005.

\bibitem[{\textit{Goldstein}(1983)}]{Goldstein_1983}
Goldstein, H., On the field configuration in magnetic clouds, in \textit{Sol.
  Wind Five}, p. 731, NASA Conf. Publ. 2280, Washington D. C., 1983.

\bibitem[{\textit{Good and Forsyth}(2016)}]{Good_Forsyth_2016}
Good, S., and R.~Forsyth, Interplanetary coronal mass ejections observed by
  {MESSENGER} and {Venus Express}, \textit{Sol. Phys.}, \textit{291}, 239--263,
  2016.

\bibitem[{\textit{Gopalswamy et~al.}(2003)\textit{Gopalswamy, Shimojo, Lu,
  Yashiro, Shibasaki, and Howard}}]{Gopalswamy_etal_2003}
Gopalswamy, N., M.~Shimojo, W.~Lu, S.~Yashiro, K.~Shibasaki, and R.~A. Howard,
  Prominence eruptions and coronal mass ejection: A statistical study using
  microwave observations, \textit{Astrophys. J.}, \textit{586}, 562--578, 2003.

\bibitem[{\textit{Gopalswamy et~al.}(2009)\textit{Gopalswamy, M{\"a}kel{\"a},
  Xie, Akiyama, and Yashiro}}]{Gopalswamy_etal_2009}
Gopalswamy, N., P.~M{\"a}kel{\"a}, H.~Xie, S.~Akiyama, and S.~Yashiro, {CME}
  interactions with coronal holes and their interplanetary consequences,
  \textit{J. Geophys. Res.}, \textit{114}, A00A22, 2009.

\bibitem[{\textit{Gui et~al.}(2011)\textit{Gui, Shen, Wang, Ye, Liu, Wang, and
  Zhao}}]{Gui_etal_2011}
Gui, B., C.~Shen, Y.~Wang, P.~Ye, J.~Liu, S.~Wang, and X.~Zhao, Quantitative
  analysis of cme deflections in the corona, \textit{Sol. Phys.}, \textit{271},
  111--139, 2011.

\bibitem[{\textit{Hidalgo and
  Nieves-Chinchilla}(2012)}]{Hidalgo_Nieves-Chinchilla_2012}
Hidalgo, M.~A., and T.~Nieves-Chinchilla, A global magnetic topology model for
  magnetic clouds. i., \textit{Astrophys. J.}, \textit{748}, 109(7pp), 2012.

\bibitem[{\textit{Hidalgo et~al.}(2002)\textit{Hidalgo, Nieves-Chinchilla, and
  Cid}}]{Hidalgo_etal_2002a}
Hidalgo, M.~A., T.~Nieves-Chinchilla, and C.~Cid, Elliptical cross-section
  model for the magnetic topology of magnetic clouds, \textit{Geophys. Res.
  Lett.}, \textit{29}, 1637, 2002.

\bibitem[{\textit{Hoeksema et~al.}(2014)}]{Hoeksema_etal_2014}
Hoeksema, J.~T., et~al., The helioseismic and magnetic imager {(HMI)} vector
  magnetic field pipeline: Overview and performance, \textit{Sol. Phys.},
  \textit{289}, 3483--3530, 2014.

\bibitem[{\textit{Howard and Harrison}(2013)}]{HowardT_Harrison_2013}
Howard, T.~A., and R.~A. Harrison, Stealth coronal mass ejections: A
  perspective, \textit{Sol. Phys.}, \textit{285}, 269--280, 2013.

\bibitem[{\textit{Hu and Sonnerup}(2002)}]{Hu_Sonnerup_2002}
Hu, Q., and B.~U.~O. Sonnerup, Reconstruction of magnetic clouds in the solar
  wind: Orientations and configurations, \textit{J. Geophys. Res.},
  \textit{107}, 1142, 2002.

\bibitem[{\textit{Hu et~al.}(2004)\textit{Hu, Smith, Ness, and
  Skoug}}]{Hu_etal_2004}
Hu, Q., C.~W. Smith, N.~F. Ness, and R.~M. Skoug, Multiple flux rope magnetic
  ejecta in the solar wind, \textit{J. Geophys. Res.}, \textit{109(A3)},
  A03,102, 2004.

\bibitem[{\textit{Isavnin et~al.}(2013)\textit{Isavnin, Vourlidas, and
  Kilpua}}]{Isavnin_etal_2013}
Isavnin, A., A.~Vourlidas, and E.~K.~J. Kilpua, Three-dimensional evolution of
  erupted flux ropes from the {Sun} (2--20 r$\odot$) to 1 {AU}, \textit{Sol.
  Phys.}, \textit{284}, 203--215, 2013.

\bibitem[{\textit{Isavnin et~al.}(2014)\textit{Isavnin, Vourlidas, and
  Kilpua}}]{Isavnin_etal_2014}
Isavnin, A., A.~Vourlidas, and E.~K.~J. Kilpua, Three-dimensional evolution of
  flux-rope {CMEs} and its relation to the local orientation of the
  heliospheric current sheet, \textit{Sol. Phys.}, \textit{289}, 2141--2156,
  2014.

\bibitem[{\textit{Janvier et~al.}(2013)\textit{Janvier, Demoulin, and
  Dasso}}]{Janvier_etal_2013}
Janvier, M., P.~D\'{e}moulin, and S.~Dasso, Global axis shape of magnetic clouds
deduced from the distribution of their local axis orientation, \textit{Astron. \& Astrophys.}, \textit{556}, A50,
  2013.

\bibitem[{\textit{Kaiser et~al.}(2008)\textit{Kaiser, Kucera, Davila, {St.
  Cyr}, Guhathakurta, and Christian}}]{Kaiser_etal_2008}
Kaiser, M.~L., T.~A. Kucera, J.~M. Davila, O.~C. {St. Cyr}, M.~Guhathakurta,
  and E.~Christian, The stereo mission: An introduction, \textit{Space Sci.
  Rev.}, \textit{136}, 5--16, 2008.

\bibitem[{\textit{Kataoka et~al.}(2015)\textit{Kataoka, Shiota, Kilpua, and
  Keika}}]{Kataoka_etal_2015}
Kataoka, R., D.~Shiota, E.~Kilpua, and K.~Keika, Pileup accident hypothesis of
  magnetic storm on 17 {March} 2015, \textit{Geophys. Res. Lett.}, \textit{42},
  5155--5161, 2015.

\bibitem[{\textit{Kay and Opher}(2015)}]{Kay_Opher_2015}
Kay, C., and M.~Opher, The heliocentric distance where the deflections and
  rotations of solar coronal mass ejections occur, \textit{Astrophys. J.
  Lett.}, \textit{811}, L36(6pp), 2015.

\bibitem[{\textit{Kay et~al.}(2013)\textit{Kay, Opher, and
  Evans}}]{Kay_etal_2013}
Kay, C., M.~Opher, and R.~M. Evans, Forecasting a coronal mass ejection's
  altered trajectory: {ForeCAT}, \textit{Astrophys. J.}, \textit{775}, 5(17pp),
  2013.

\bibitem[{\textit{Kilpua et~al.}(2009)\textit{Kilpua, Pomoell, Vourlidas,
  Vainio, Luhmann, Li, Schroeder, Galvin, and Simunac}}]{Kilpua_etal_2009}
Kilpua, E. K.~J., J.~Pomoell, A.~Vourlidas, R.~Vainio, J.~Luhmann, Y.~Li,
  P.~Schroeder, A.~B. Galvin, and K.~Simunac, {STEREO} observations of
  interplanetary coronal mass ejections and prominence deflection during solar
  minimum period, \textit{Ann. Geophys.}, \textit{27}, 4491--4503, 2009.

\bibitem[{\textit{Lemen et~al.}(2012)}]{Lemen_etal_2012}
Lemen, J.~R., et~al., The atmospheric imaging assembly ({AIA}) on the solar
  dynamics observatory ({SDO}), \textit{Sol. Phys.}, \textit{275}, 17--40,
  2012.

\bibitem[{\textit{Lepping et~al.}(1990)\textit{Lepping, Jones, and
  Burlaga}}]{Lepping_etal_1990}
Lepping, R.~P., J.~A. Jones, and L.~F. Burlaga, Magnetic field structure of
  interplanetary magnetic clouds at 1 {AU}, \textit{J. Geophys. Res.},
  \textit{95}, 11,957--11,965, 1990.

\bibitem[{\textit{Lepping et~al.}(1995)}]{Lepping_etal_1995}
Lepping, R.~P., et~al., The {Wind} magnetic field investigation, \textit{Space
  Sci. Rev.}, \textit{71}, 207--229, 1995.

\bibitem[{\textit{Lin et~al.}(1995)}]{Lin_etal_1995}
Lin, R.~P., et~al., A three-dimensional plasma and energetic particle
  investigation for the {Wind} spacecraft, \textit{Space Sci. Rev.},
  \textit{71}, 125--153, 1995.

\bibitem[{\textit{Liu et~al.}(2015)\textit{Liu, Hu, Wang, Yang, Zhu, Liu,
  Luhmann, and Richardson}}]{LiuY_etal_2015}
Liu, Y.~D., H.~Hu, R.~Wang, Z.~Yang, B.~Zhu, Y.~A. Liu, J.~G. Luhmann, and
  J.~D. Richardson, Plasma and magnetic field characteristics of solar coronal
  mass ejections in relation to geomangetic strom intensity and variability,
  \textit{Astrophys. J. Lett.}, \textit{809}, L34(6pp), 2015.

\bibitem[{\textit{Lopez and Freeman}(1986)}]{Lopez_Freeman_1986}
Lopez, R.~E., and J.~W. Freeman, Solar wind proton temperature-velocity
  relationship, \textit{J. Geophys. Res.}, \textit{91}, 1701, 1986.

\bibitem[{\textit{Lugaz et~al.}(2010)\textit{Lugaz, {Hernandez-Charpak},
  Roussev, Davis, Vourlidas, and Davies}}]{Lugaz_etal_2010}
Lugaz, N., J.~N. {Hernandez-Charpak}, I.~I. Roussev, C.~J. Davis, A.~Vourlidas,
  and J.~A. Davies, Determining the azimuthal properties of coronal mass
  ejections from multi-spacecraft remote-sensing observations with {STEREO
  SECCHI}, \textit{Astrophys. J.}, \textit{715}, 493--499, 2010.

\bibitem[{\textit{MacQueen et~al.}(1986)\textit{MacQueen, Hundhausen, and
  Conover}}]{MacQueen_etal_1986}
MacQueen, R.~M., A.~J. Hundhausen, and C.~W. Conover, The propagation of
  coronal mass ejection transients, \textit{J. Geophys. Res.}, \textit{91},
  31--38, 1986.

\bibitem[{\textit{Marubashi}(1986)}]{Marubashi_1986}
Marubashi, K., Structure of the interplanetary magnetic clouds and their solar
  origins, \textit{Adv. in Space Res.}, \textit{6}, 335--338, 1986.

\bibitem[{\textit{Marubashi and Lepping}(2007)}]{Marubashi_Lepping_2007}
Marubashi, K., and R.~P. Lepping, Long-duration magnetic clouds: a comparison
  of analyses using torus- and cylinder-shaped flux rope models, \textit{Ann.
  Geophys.}, \textit{25}, 2453--2477, 2007.

\bibitem[{\textit{M{\"o}stl et~al.}(2015)}]{Mostl_etal_2015}
M{\"o}stl, C., et~al., Strong coronal channelling and interplanetary evolution
  of a solar storm up to {Earth} and {Mars}, \textit{Nature Commun.},
  \textit{6}, 7135, 2015.

\bibitem[{\textit{Mulligan and Russell}(2001)}]{Mulligan_Russell_2001}
Mulligan, T., and C.~T. Russell, Multispacecraft modeling of the flux rope
  structure of interplanetary coronal mass ejections: Cylindrically symmetric
  versus nonsymmetric topologies, \textit{J. Geophys. Res.}, \textit{106(A6)},
  10,581--10,596, 2001.

\bibitem[{\textit{Mulligan et~al.}(1998)\textit{Mulligan, Russell, and
  Luhmann}}]{Mulligan_etal_1998}
Mulligan, T., C.~T. Russell, and J.~G. Luhmann, Solar cycle evolution of the
  structure of magnetic clouds in the inner heliosphere, \textit{Geophys. Res.
  Lett.}, \textit{25}, 2959--2962, 1998.

\bibitem[{\textit{Ogilvie et~al.}(1995)}]{Ogilvie_etal_1995}
Ogilvie, K.~W., et~al., {SWE}, a comprehensive plasma instrument for the {Wind}
  spacecraft, \textit{Space Sci. Rev.}, \textit{71}, 55--77, 1995.

\bibitem[{\textit{Riley et~al.}(2004)}]{Riley_etal_2004}
Riley, P., et~al., Fitting flux ropes to a global {MHD} solution: a comparison
  of techniques, \textit{J. Atmos. Solar-Terres. Phys.}, \textit{66},
  1321--1331, 2004.

\bibitem[{\textit{Robbrecht et~al.}(2009)\textit{Robbrecht, Patsourakos, and
  Vourlidas}}]{Robbrecht_etal_2009}
Robbrecht, E., S.~Patsourakos, and A.~Vourlidas, No trace left behind: {STEREO}
  observation of a coronal mass ejection without low coronal signatures,
  \textit{Astrophys. J.}, \textit{701}, 283--291, 2009.

\bibitem[{\textit{Romashets and Vandas}(2003)}]{Romashets_Vandas_2003}
Romashets, E.~P., and M.~Vandas, Force-free field inside a toroidal magnetic
  cloud, \textit{Geophys. Res. Lett.}, \textit{30}, 2065, 2003.

\bibitem[{\textit{Rust and Kumar}(1996)}]{Rust_Kumar_1996}
Rust, K., and A.~Kumar, Evidence for helical kinked magnetic flux ropes in
  solar eruptions, \textit{Astrophys. J. Lett.}, \textit{464}, L199--L202,
  1996.

\bibitem[{\textit{Shen et~al.}(2011{\natexlab{a}})\textit{Shen, Wang, Gui, Ye,
  and Wang}}]{Shen_etal_2011}
Shen, C., Y.~Wang, B.~Gui, P.~Ye, and S.~Wang, Kinematic evolution of a slow
  {CME} in near solar space viewed by {STEREO-B} in {October} 8, 2007,
  \textit{Sol. Phys.}, \textit{269}, 389--400, 2011{\natexlab{a}}.

\bibitem[{\textit{Shen et~al.}(2009)\textit{Shen, Feng, and
  Song}}]{ShenF_etal_2009}
Shen, F., X.~Feng, and W.~B. Song, An asynchronous and parallel time-marching
  method: application to the three-dimensional {MHD} simulation of the solar
  wind, \textit{Science in china Series E: Technological Sciences},
  \textit{52}, 2895--2902, 2009.

\bibitem[{\textit{Shen et~al.}(2011{\natexlab{b}})\textit{Shen, Feng, Wang, Wu,
  Song, Guo, and Zhou}}]{ShenF_etal_2011a}
Shen, F., X.~S. Feng, Y.~Wang, S.~T. Wu, W.~B. Song, J.~P. Guo, and Y.~F. Zhou,
  Three-dimensional {MHD} simulation of two coronal mass ejections' propagation
  and interaction using a successive magnetized plasma blobs model, \textit{J.
  Geophys. Res.}, \textit{116}, A09,103, 2011{\natexlab{b}}.

\bibitem[{\textit{Thernisien}(2011)}]{Thernisien_2011}
Thernisien, A., Implementation of the graduated cylindrical shell model for the
  three-dimensional reconstruction of coronal mass ejections,
  \textit{Astrophys. J. Suppl. Ser.}, \textit{194}, 33, 2011.

\bibitem[{\textit{Thernisien et~al.}(2009)\textit{Thernisien, Vourlidas, and
  Howard}}]{Thernisien_etal_2009}
Thernisien, A., A.~Vourlidas, and R.~Howard, Forward modeling of coronal mass
  ejections using {STEREO/SECCHI} data, \textit{Sol. Phys.}, \textit{256},
  111--130, 2009.

\bibitem[{\textit{Vandas and Romashets}(2003)}]{Vandas_Romashets_2003}
Vandas, M., and E.~P. Romashets, A force-free field with constant alpha in an
  oblate cylinder: A generalization of the lundquist solution, \textit{Astron.
  \& Astrophys.}, \textit{398}, 801--807, 2003.

\bibitem[{\textit{Vr\v{s}nak et~al.}(2010)\textit{Vr\v{s}nak, \v{Z}ic,
  Falkenberg, M{\"o}stl, Vennerstrom, and Vrbanec}}]{Vrsnak_etal_2010}
Vr\v{s}nak, B., T.~\v{Z}ic, T.~V. Falkenberg, C.~M{\"o}stl, S.~Vennerstrom, and
  D.~Vrbanec, The role of aerodynamic drag in propagation of interplanetary
  coronal mass ejections, \textit{Astron. \& Astrophys.}, \textit{512}, A43,
  2010.

\bibitem[{\textit{Vr\v{s}nak et~al.}(2013)}]{Vrsnak_etal_2013}
Vr\v{s}nak, B., et~al., Propagation of interplanetary coronal mass ejections:
  The drag-based model, \textit{Sol. Phys.}, \textit{285}, 295--315, 2013.

\bibitem[{\textit{Wang et~al.}(2002)\textit{Wang, Ye, Wang, Zhou, and
  Wang}}]{Wang_etal_2002a}
Wang, Y., P.~Z. Ye, S.~Wang, G.~P. Zhou, and J.~X. Wang, A statistical study on
  the geoeffectiveness of earth-directed coronal mass ejections from {March}
  1997 to {December} 2000, \textit{J. Geophys. Res.}, \textit{107(A11)}, 1340,
  doi:10.1029/2002JA009,244, 2002.

\bibitem[{\textit{Wang et~al.}(2004)\textit{Wang, Shen, Ye, and
  Wang}}]{Wang_etal_2004b}
Wang, Y., C.~Shen, P.~Ye, and S.~Wang, Deflection of coronal mass ejection in
  the interplanetary medium, \textit{Sol. Phys.}, \textit{222}, 329--343, 2004.

\bibitem[{\textit{Wang et~al.}(2006)\textit{Wang, Xue, Shen, Ye, Wang, and
  Zhang}}]{Wang_etal_2006a}
Wang, Y., X.~Xue, C.~Shen, P.~Ye, S.~Wang, and J.~Zhang, Impact of the major
  coronal mass ejections on geo-space during {September} 7 -- 13, 2005,
  \textit{Astrophys. J.}, \textit{646}, 625--633, 2006.

\bibitem[{\textit{Wang et~al.}(2011)\textit{Wang, Chen, Gui, Shen, Ye, and
  Wang}}]{Wang_etal_2011}
Wang, Y., C.~Chen, B.~Gui, C.~Shen, P.~Ye, and S.~Wang, Statistical study of
  coronal mass ejection source locations: Understanding cmes viewed in
  coronagraphs, \textit{J. Geophys. Res.}, \textit{116}, A04,104,
  doi:{10.1029/2010JA016,101}, 2011.

\bibitem[{\textit{Wang et~al.}(2014)\textit{Wang, Wang, Shen, Shen, and
  Lugaz}}]{Wang_etal_2014}
Wang, Y., B.~Wang, C.~Shen, F.~Shen, and N.~Lugaz, Deflected propagation of a
  coronal mass ejection from the corona to interplanetary space, \textit{J.
  Geophys. Res.}, \textit{119}, 5117--5132, 2014.

\bibitem[{\textit{Wang et~al.}(2015)\textit{Wang, Zhou, Shen, Liu, and
  Wang}}]{Wang_etal_2015}
Wang, Y., Z.~Zhou, C.~Shen, R.~Liu, and S.~Wang, Investigating plasma motion of
  magnetic clouds at 1 {AU} through a velocity-modified cylindrical force-free
  flux rope model, \textit{J. Geophys. Res.}, \textit{120}, 1543--1565, 2015.

\bibitem[{\textit{Webb et~al.}(2001)\textit{Webb, Crooker, Plunkett, and {St.
  Cyr}}}]{Webb_etal_2001}
Webb, D.~F., N.~U. Crooker, S.~P. Plunkett, and O.~C. {St. Cyr}, The solar
  sources of geoeffective structures, in \textit{Space Weather}, edited by
  S.~Paul, J.~S. Howard, and L.~S. George, Geophys. Monogr. Ser. 125, pp.
  123--142, AGU, 2001.

\bibitem[{\textit{Wood et~al.}(2016)\textit{Wood, Lean, McDonald, and
  Wang}}]{Wood_etal_2016}
Wood, B.~E., J.~L. Lean, S.~E. McDonald, and Y.-M. Wang, Comparative
  ionospheric impacts and solar origins of nine strong geomagnetic storms in
  2010-2015, \textit{J. Geophys. Res.}, \textit{accepted},
  doi:10.1002/2015JA021,953, 2016.

\bibitem[{\textit{Yermolaev et~al.}(2005)\textit{Yermolaev, Yermolaev,
  Zastenker, Zelenyi, Petrukovich, and Sauvaud}}]{Yermolaev_etal_2005}
Yermolaev, Y.~I., M.~Yermolaev, G.~Zastenker, L.~Zelenyi, A.~Petrukovich, and
  J.-A. Sauvaud, Statistical studies of geomagnetic storm dependencies on solar
  and interplanetary events: a review, \textit{Planet. Space Sci.},
  \textit{53}, 189--196, 2005.

\bibitem[{\textit{Zhang et~al.}(2007)}]{Zhang_etal_2007}
Zhang, J., et~al., Solar and interplanetary sources of major geomagnetic storms
  ({Dst} $\leq$ -100 {nT}) during 1996-2005, \textit{J. Geophys. Res.},
  \textit{112}, A10,102, 2007.

\bibitem[{\textit{Zhao and Webb}(2003)}]{Zhao_Webb_2003}
Zhao, X.~P., and D.~F. Webb, Source regions and storm effectiveness of
  frontside full halo coronal mass ejections, \textit{J. Geophys. Res.},
  \textit{108(A6)}, 1234, 2003.

\bibitem[{\textit{Zuccarello et~al.}(2012)\textit{Zuccarello, Bemporad, Jacobs,
  Mierla, Poedts, and Zuccarello}}]{Zuccarello_etal_2012}
Zuccarello, F.~P., A.~Bemporad, C.~Jacobs, M.~Mierla, S.~Poedts, and
  F.~Zuccarello, The role of streamers in the deflection of coronal mass
  ejections: Comparison between {STEREO} three-dimensional reconstructions and
  numerical simulations, \textit{Astrophys. J.}, \textit{744}, 66(14pp), 2012.

\end{thebibliography}

\end{article}
\end{document}